\documentstyle[preprint,aps]{revtex}
\tighten
\begin{document}                               
\draft
\hyphenation{Rijken}
\hyphenation{Nijmegen}
\hyphenation{Stich-ting Fun-da-men-teel On-der-zoek Ma-te-rie}
\hyphenation{Ne-der-land-se Or-ga-ni-sa-tie We-ten-schap-pe-lijk}

\title{Triton calculations with the new Nijmegen potentials}

\author{J.L.\ Friar}
\address{Theoretical Division, Los Alamos National Laboratory,
         Los Alamos, NM 87545 \\
         and
         Institut f\"ur Kernphysik, Johannes Gutenberg-Universit\"at,
         D-6500 Mainz, Germany}

\author{G.L.\ Payne}
\address{Department of Physics and Astronomy, University of Iowa,
         Iowa City, IA 52242}

\author{V.G.J.\ Stoks and J.J.\ de Swart}
\address{Institute for Theoretical Physics, University of Nijmegen,
         Nijmegen, The Netherlands}

\date{submitted to Phys.\ Lett.\ B}
\maketitle

\begin{abstract}
Triton properties are calculated using new nucleon-nucleon
potentials, which were fit to the world nucleon-nucleon data.
All potentials are charge dependent and explicitly incorporate
the mass difference between the charged and neutral pions.
Three of these models have a nearly optimal $\chi^{2}$ per degree
of freedom and can therefore be considered as alternative
partial-wave analyses, which in quality can almost compete with
the Nijmegen partial-wave analysis.
The triton binding energy obtained with three local models
(Nijm~II, Reid93, AV18) can be summarized as 7.62$\pm$0.01 MeV,
which is nearly 900 keV lower than experiment.
The non-local model Nijm~I binds by 7.72 MeV.
\end{abstract}

\narrowtext

\section{FEW-NUCLEON SYSTEMS}
In recent years one of the important problems of few-nucleon
physics has been resolved. The non-relativistic Schr\"{o}dinger
(or Faddeev~\cite{Fad60}) equation can now be solved numerically
for an arbitrary (energy-independent) potential model, and with
negligible error. Such complete or ``exact'' solutions have been
carried out at the level of 1\% (or less) error for the ground
states~\cite{Che85} of $^3$He and $^3$H (including a Coulomb
interaction in the former case), for the ground state~\cite{Car87}
of $^4$He, for the low-lying (continuum) states~\cite{Carunp} of
$^5$He, for the zero-energy scattering states of the $n$-$d$ and
$p$-$d$ systems~\cite{Che91} and for transitions~\cite{Fri91}
between these systems and the ground states, and for the $n$-$d$
continuum states above breakup threshold~\cite{Glo90}.

The results of these calculations can be rather simply summarized.
The ground states are underbound and their underbinding is correlated.
The triton is underbound by an amount which varies from 0.2 to
1.1 MeV for a wide variety of potential models~\cite{Che85}.
The sizes of these systems are strongly correlated with their
binding, and reasonable extrapolations to the physical binding
energies of $^3$H and $^3$He produce sizes which agree with
experiment.
The Coulomb energy~\cite{Fad60} of $^3$He (properly extrapolated)
accounts for about 85\% of the $^3$He--$^3$H binding-energy
difference, and a variety of small charge-symmetry-breaking
mechanisms~\cite{Wu90} produces the remaining 15\%.
The alpha particle is underbound by an amount which is correlated
with the triton underbinding (by roughly a factor of four) and
correcting one problem will likely correct the other. Low-energy
$n$-$d$ and $p$-$d$ scattering and capture reactions are in
reasonable agreement~\cite{Che91,Fri91} with experiment (after
extrapolating), except for the $p$-$d$ scattering length, whose
experimental value may be suspect.
Scattering calculations (above breakup threshold~\cite{Glo90})
are almost entirely in agreement with experiment. Those few areas
where disagreements exist could be the result of using rather
poor two-body forces, comparing $n$-$d$ calculations to
(Coulomb-modified) $p$-$d$ data, inaccurate data, or an
incompletely understood nuclear force. Few groups have performed
these difficult scattering calculations.

Among the interesting physics which might manifest itself in these
comparisons is three-nucleon forces~\cite{Coo79}. Traditionally, the
nuclear force is described by a sum of two-body (pairwise) potentials.
However, more complicated three-body forces are present; the latter
require the simultaneous specification of the spatial coordinates of
all three nucleons, as well as their spin and isospin states.
These three-nucleon forces are expected to contribute a rather small
amount of binding, which can be understood as follows.
Their amount of binding is schematically given by
$\langle V_{\pi}\rangle ^2/Mc^2$, where $\langle V_{\pi}\rangle$
is the contribution of OPEP (the one-pion-exchange potential) to the
triton potential energy and $M$ is the nucleon mass, which has been
chosen to be the generic ``large mass'' scale which remains after one
treats the pion. Using $\langle V_{\pi}\rangle\sim$ 30--40 MeV, one
estimates the contribution of three-nucleon forces to be 1--2 MeV
to the triton potential energy (out of a total of roughly 50 MeV).
Considerable effort has been made to calculate the longest-range part
of the three-nucleon force (due to two-pion exchange). However, the
short-range part of that force is not easily amenable to theoretical
treatment and unfortunately the triton binding is sensitive to that
short-range behavior.
Calculations of the triton binding which include three-nucleon
forces are consistent~\cite{Fad60} with the estimates given above.

The scale of a few percent of the kinetic or potential energy is
also the scale for relativistic corrections. A representative momentum
scale in the few-nucleon systems is the pion mass: $m_{\pi}c$.
An estimate for the size of the relativistic corrections for the
nucleons is then $(v/c)^2 \sim(m_{\pi}/M)^2 \sim 2\%$,
which is indeed the size of correction we estimated for three-body
forces. Calculations performed to date typically find corrections
of small magnitude from special relativity ($\sim$ 0.2--0.3 MeV),
but are not otherwise in agreement~\cite{Car93,Rup92,Glo86,Sam92}.
Complicating matters further is the fact that at least a subset
of relativistic corrections is comprised of three-nucleon forces.

In addition to the uncertainties from relativity and three-body-force
mechanisms there is the uncertainty in the two-nucleon
force. All of the calculations described above were performed
with two-nucleon forces which are called ``realistic.''
Such forces contain OPEP and give at least a qualitative fit to
the scattering data. The $\chi^2$ per degree of freedom for such
a potential compared to the scattering data can nevertheless vary
greatly. This is particularly true when a potential which has been
fit to the $np$ data is compared to the $pp$ data~\cite{St93a}.
One of the reasons being that charge dependence between the $T=1$
$np$ and $pp$ partial waves in such a fit is not always accounted
for properly (e.g., effects due to the presence of the Coulomb
interaction are included, whereas equally important effects due to
the neutral-to-charged pion mass difference are completely neglected).
Another reason is that the $np$ data are not nearly as
accurate as the $pp$ data. So in a fit to only the $np$ data the
constraints on the parameters of the model are less restrictive,
which can result in a set of parameters which give a poor description
of the $pp$ data.
The parameters should be fit to both $pp$ and $np$ data.
Differences in the quality of the fit to the data will produce
differences in the predicted triton binding. In addition, differences
in the assumed form of the potential, or of its type (local, or
momentum dependent), might also affect the binding
energy and yet not affect the fit to the scattering data in the
0--350 MeV energy region, which is the traditional domain of
nucleon-nucleon ($N\!N$) potential models.

While it is not yet possible to resolve problems with the
three-nucleon forces and with relativistic corrections, it should
be possible to eliminate much of the uncertainty associated with
the two-nucleon force.
In the triton the charge dependence of the nuclear force can be taken
into account by using an effective charge-symmetric $^{1}S_{0}$
force given by~\cite{Fri87}
\begin{equation}
   V_{\rm eff} = \frac{2}{3}\, V_{pp} + \frac{1}{3}\, V_{np} \ ,
\end{equation}
which prescription has an error the order of a few keV.
Potential models whose $^{1}S_{0}$
forces are fit only to $pp$ scattering generate a triton binding
energy approximately 100 keV too low, while those fit only to $np$
scattering are roughly 200 keV too high. Obviously, using a force
model which builds in the proper amount of charge dependence (i.e.,
fits both $pp$ and $np$ data) eliminates this problem.
After applying this correction to previous triton calculations,
potential models still underbind the triton by 0.4 to 1.0 MeV.
This spread of values is larger than most estimates of relativistic
effects and comparable to the contribution of most three-nucleon
forces.

It has been known for many years that weakening the $T=0$ tensor
force (but still maintaining a fit to the two-nucleon data) increases
the triton binding energy. The reason is that the deuteron is even
more sensitive to the tensor force than is the triton, in spite of
the fact that more than half of the triton's potential energy has that
source.
Consequently, when one is fitting the $N\!N$ potential to
the two-nucleon scattering data, slight variations in the tensor
force (increases or decreases) must be compensated by opposite
variations (decreases or increases) in the central force. The triton
is relatively more sensitive to the central force.
Thus a weaker tensor force increases the triton binding because
the deuteron binding is {\it fixed}.
Until recently, published measures of the tensor force were of
poor quality, and almost any force seemed acceptable. This situation
has definitely been changed with the completion of a new and
comprehensive partial-wave analysis (PWA), which we now discuss.

\section{NIJMEGEN PARTIAL-WAVE ANALYSIS \protect\\ AND POTENTIAL}
The Nijmegen partial-wave analysis has been described in detail
elsewhere~\cite{Ber88,Ber90,Klo92,St93b}. We list here a few of the
salient features which are relevant to our present calculation.

The Nijmegen procedure treats OPEP explicitly (allowing the
pion-nucleon couplings to be determined by the data) and
incorporates Coulomb, magnetic moment, and vacuum polarization
interactions explicitly in the appropriate places.
As a consistency check~\cite{Klo91} the pion masses in OPEP are allowed
to vary and the masses of the neutral and charged pions are found to
be $m_{\pi^0}=135.6(13)$ MeV and $m_{\pi^\pm}=139.4(10)$ MeV.
These results agree with the free pion masses and the small error
bars emphasize the importance of OPEP.
The pion-nucleon coupling constants are consistent with being equal.
The explicit inclusion of these different pion masses (next to
other, less important, contributions) produces a charge-dependent
nuclear force.

By fitting all the $N\!N$ scattering data simultaneously in an
energy-dependent (or multi-energy) partial-wave analysis, uncertainties
associated with poor quality data in a few kinematical regions can be
compensated by good quality data at other energies.
Energy-independent (or single-energy) partial-wave analyses do not
have this advantage. Moreover, the energy bin analyzed in a
single-energy analysis may not contain the appropriate types of
scattering data to pin down a particular phase parameter.
An example in this context is the lack of $np$ spin-correlation data
near 100 MeV, which dictates that the $J=1$ mixing parameter,
$\varepsilon_1$, cannot be determined accurately in a single-energy
analysis at this energy.
Similarly, recent data at 67 MeV~\cite{Ham91,Haf92} and their
inclusion in the partial-wave analysis~\cite{Klo91} have
helped to reduce the uncertainties associated with $\varepsilon_1$
in the single-energy analysis at 50 MeV.
On the other hand, the errors associated with a complete multi-energy
partial-wave analysis are typically much smaller than those
associated with single-energy partial-wave analyses.
The reason, of course, being that in a single-energy analysis of
one particular energy bin the information about the overall energy
dependence of the phase parameters is not incorporated.
So the set of single-energy analyses covering the typical 0--350 MeV
energy range requires many more fit parameters than the multi-energy
analysis covering the same energy range.
In the multi-energy analysis the $\varepsilon_1$ mixing parameter
can be determined very well {\it at all energies}, due to the presence
of spin-correlation data at various different energies throughout the
0--350 MeV region~\cite{Klo91}.
This is demonstrated clearly in Fig.~\ref{eps1}, which depicts the
Nijmegen multi-energy and single-energy values and errors for that
quantity~\cite{St93b}.

Since $\varepsilon_1$ is the most commonly used measure of the
tensor force, it is incorrect to state that virtually any tensor
force is consistent with the $N\!N$ scattering data.
Figure~\ref{eps1} demonstrates unequivocally that there are very
tight constraints on the tensor force. This is also illustrated in
Table~\ref{results}, where we give $\varepsilon_1$ at 50 MeV of the
Nijmegen multi-energy partial-wave analysis~\cite{St93b} (Nijm PWA93)
and of the various new potential models to be discussed next.
The spread in $\varepsilon_1$ values of the different models
(Nijm PWA93, Nijm~I, Nijm~II, and Reid93) gives an indication for the
systematic error on $\varepsilon_1$, which we believe is about
0.05$^{\circ}$. However, one has to bear in mind that the potential
models, though of high quality, are still not as good as the
Nijm PWA93 analysis.

The original Nijmegen potential~\cite{Nag78} (Nijm78) is a
one-boson-exchange potential which incorporates the non-strange
mesons of the pseudoscalar, vector, and scalar nonets. These
mesons can be identified with the lowest-lying Regge trajectories.
The identification leads to a Gaussian regularization of the
short-range behavior. Using 13 parameters, the description of the
$pp$ data is reasonably good~\cite{St93a}, whereas the description of
the $np$ data is rather poor. In order to improve its quality, we are
currently constructing an update using 15 parameters. The preliminary
version (Nijm92) gives a much better description of the $pp$ as well as
the $np$ data with $\chi^{2}$ per datum of 1.92. However, its quality
is still not as good as the quality of the Nijmegen partial-wave
analysis Nijm PWA93, which has $\chi^{2}$ per datum of 0.99.
We therefore also followed a different approach in that we constructed
a Reid-like model where each partial wave is parametrized independently.
Introducing as many parameters as necessary, it is then easy to arrive
at a model with a (nearly) optimal $\chi^{2}$ per datum. The Reid-like
Nijmegen model constructed this way is denoted by Nijm~I.

A feature of relativistic origin in the Nijmegen potentials is the
momentum-dependent part of the central potential, which follows from
field theory. It gives rise to a non-local structure
$(\Delta \phi(r)+\phi(r)\Delta)$ to the potential in
configuration space (see, e.g., Ref.~\cite{Nag78}). Such a term might
be expected to behave differently in the triton, which is one of the
reasons we constructed two Reid-like versions of the Nijmegen potential:
The non-local Nijm~I potential, which contains these momentum-dependent
terms (as do the Nijm78 and Nijm92 potentials), and a local Nijm~II
potential, where these terms are intentionally omitted.
We also constructed updates of two other {\it local\/} potentials.
The Reid soft-core potential~\cite{RSC68} was reparametrized using
sums of regularized Yukawa functions and is here denoted by Reid93.
The Argonne potential~\cite{Wir84} was extended to include
charge-independence breaking in the phenomenological parametrization
of the short-range interaction and is here denoted by AV18. All these
models will be discussed in more detail elsewhere~\cite{St93c}.
Here we only want to mention that all potentials explicitly include the
charge-dependent OPEP described earlier.

It is important to note that the Nijmegen partial-wave analysis Nijm
PWA93 has not been used in constructing these potentials. The parameters
of all models have been optimized in a direct fit to the data. In fact,
both the Nijmegen potentials Nijm~I and Nijm~II, as well as the
regularized Reid soft-core potential Reid93, are alternative
partial-wave analyses: They have roughly the same number of fit
parameters as our original partial-wave analysis Nijm PWA93, they are
fit to the same data, and they achieve nearly the same values of
$\chi^2_{\rm min}$, which is significantly better than any
other previous potential model. It is even significantly better than
other multi-energy partial-wave analyses.
The $\chi^2_{\rm min}$ per datum for the new models is given in
Table~\ref{results}.

\section{\bf TRITON CALCULATIONS}
The potentials used in this work comprise two Reid-like Nijmegen models
(Nijm~I and Nijm~II), a regularized update of the Reid soft-core
model (Reid93), and preliminary updates of the Nijmegen and Argonne
models (Nijm92 and AV18, respectively).
The Faddeev equations for the triton bound state were solved
for these new potentials using up to 34 channels (three-nucleon
partial-wave states), which guarantees partial-wave convergence
at the level of 10 keV. This includes all partial waves of the
$N\!N$ interaction with $J\leq4$.

We should also mention that the new potential models have been fit
to the deuteron binding energy using relativistic kinematics, which
means that the binding energy of the deuteron is taken to be
$2M-2\sqrt{M^2-\kappa^2}$, rather than $\kappa^2/M$.
The difference is very small, and versions of the Nijm~I and Nijm~II
potentials were constructed to accommodate the latter form; this changes
the triton binding by less than 1.5 keV, a truly negligible effect.

The triton bound-state result for the non-local Nijm~I potential for
34 channels is 7.72 MeV, which is nearly 800 keV lower than the
experimental value of 8.48 MeV.
The charge radii for $^3$H and $^3$He, and the Coulomb energy of
$^3$He are not significantly different from what one would expect
for these binding energies.

The results for this Nijm~I potential are not significantly different
from those of the original Nijm78 potential, whose binding energy for
the triton is 7.63 MeV. That potential, however, had a $pp$-type
$^{1}S_{0}$ force, so that the result should be increased by roughly
100 keV. We note that these similar results should not be unexpected
in view of a rather similar tensor force in the two models. The Nijm78
and Nijm~I models have deuteron D-state probabilities of 5.39 and
5.66\%, respectively. Moreover, the two models have an identical
structural form. We also note that the Nijm~I potential generates
72\% of the total triton potential energy (48~MeV) from the tensor
force and about 74\% of that energy from the (iterated) OPEP component
(i.e., $\langle V_{\pi}\rangle$). The bulk of the potential energy
typically comes from the tensor force and OPEP.

The quality of the local Nijm~II potential is equally as good as that
of the non-local Nijm~I potential, both having a $\chi^2$ per datum of
1.03. The deuteron D-state probability is 5.64\%, which is virtually
the same as that of the Nijm~I potential. Triton calculations with
the Nijm~II potential produces 7.62 MeV for 34 channels.
This is a strong indication that replacing local structure in
configuration space by a non-local one can affect the triton
binding energy, even when the quality of the fit to the $N\!N$ data
remains unchanged. This potential generates 52\% and 67\% of the
triton potential energy from the tensor and (iterated) OPEP forces,
respectively.

A significant question which can be posed is the following:
how important are differences in the specific functional forms
used to parameterize the radial dependence in the various parts
of the potential? In order to answer this question in part, we also
solved the triton bound state for the Reid93 and AV18 potentials.
The Reid93 potential has an equally good fit to the $N\!N$ data.
The deuteron D-state probability is 5.70\%, slightly higher than
the other potentials. The triton binding energy for this potential
is 7.63 MeV, which is very close to that of the Nijm~II potential.
It is worth noting that the original Reid68 potential had a triton
binding energy of 7.35 MeV. The 300 keV difference in binding is the
result of different quality fits to the $N\!N$ data, and to a very
different current data set than existed in 1968.
The updated Argonne potential AV18 fits the $N\!N$ data with a
$\chi^{2}$ per datum of 1.30, has a D-state probability of 5.65\%,
and produces a triton binding energy of 7.62 MeV.

Finally, we mention that the (non-local) preliminary update of the
original Nijm78 potential, denoted by Nijm92, gives a triton binding
energy of 7.68 MeV. The difference with the Nijm~I result is at least
partially due to the difference in amount of non-locality.
The results for all five potential models are
given in Table~\ref{results}. Note that the Nijm92 potential has the
largest $\langle V_T \rangle$ and $\langle V_{\pi} \rangle$, indicating
a very strong tensor force. Surprisingly, its values for $P_d$ and
$\varepsilon_1$ at 50 MeV are the smallest, implying an anticorrelation.

All of the local potentials treated here have virtually the same
triton binding energy (7.62 MeV) and deuteron D-state probability
(5.66\%). This similarity may not be coincidental.
If one fits the partial waves of the potential to the partial-wave
``data'' which result from a partial-wave analysis, and if one further
assumes that the potential is local, the resulting fit to the relevant
deuteron properties and phase-shift data {\it at all energies\/} yields
a unique potential~\cite{Cha77}. In practice one does not fit to data
at all energies, of course, but it has long been an article of faith in
nuclear physics that scattering data at high energies are unimportant
to calculations of the nuclear bound states. Equivalently, these data
primarily determine details of the potential at very short distances,
which are known to be rather unimportant. Our results here are
consistent with this picture.

In general, the local potentials tend to have a fairly large binding
energy difference between the 5- and 34-channel cases ($>200$~keV),
while the opposite is true for the non-local potentials. The results for
the non-local Nijm~I potential for 5, 9, 18, 26, and 34 channels are:
7.70, 7.77, 7.67, 7.72, and 7.72 MeV. The results for the local Nijm~II
potential are 7.39, 7.56, 7.51, 7.61, and 7.62 MeV for 5, 9, 18, 26,
and 34 channels, respectively.
This trend is consistent with most other potential models~\cite{Che85}.

\section{\bf SUMMARY}
Solutions to the Faddeev equations for the triton ground state
were obtained for five new $N\!N$ potentials, three of which fit the
$N\!N$ data with a nearly optimal $\chi^2$ per datum of 1.03.
The three local potentials (Nijm~II, Reid93, and AV18) bind the triton
by 7.62$\pm$0.01 MeV. The non-local potential Nijm~I, which is of
similar quality as Nijm~II, is more bound by roughly 100 keV.
An update of the original non-local Nijm78 potential, denoted by Nijm92,
binds the triton by 7.68 MeV.
The local potential results suggest (but obviously do not prove) that
local potentials which fit the $N\!N$ scattering data very well, bind
the triton by a unique value of about 7.62 MeV. Should this prove to
be true, the physics issues in the triton problem
(besides the question of three-nucleon forces) then shift to the origin
and presence of nonlocalities in the $N\!N$ force. Indeed, the 100 keV
difference between the Nijm~I and Nijm~II models is the effect of just
such a nonlocality.
With respect to their fit to the scattering data, these
new Nijmegen potentials are the best ever constructed, and triton
binding calculations with them provide a benchmark against which
calculations with other potential models should be compared.

\acknowledgements
The work of J.L.F.\ was performed under the auspices of the U.S.\
Department of Energy, while the work of G.L.P.\ was supported in
part by the U.S.\ Department of Energy.
The work of V.G.J.S.\ was performed under auspices of the Stichting
voor Fundamenteel Onderzoek der Materie (FOM) with financial support
from the Nederlandse Organisatie voor Wetenschappelijk Onderzoek (NWO).

We would like to thank Dr.\ Th.A.\ Rijken and R.\ Klomp for many
valuable discussions, Dr.\ R.B.\ Wiringa for his help and for letting
us use the AV18 model prior to publication, and H.\ Leeb for a
stimulating summary of the ``inverse problem.''
One of us (J.L.F.) would like to thank the Alexander von
Humboldt-Stiftung for support.

\begin{table}
\caption{Potential models used in this work, the values of
         $\chi^2_{\rm min}$ per datum obtained in producing them,
         values of $\varepsilon_{1}$ in degrees at 50~MeV and
         the deuteron D-state percentages, the corresponding
         triton binding energies, and percentage contributions of
         the tensor force and OPEP to the total potential energy.
         The first entry contains the results of the Nijmegen
         partial-wave analysis~\protect\cite{St93b} (Nijm PWA93)
         for comparison.}
\begin{tabular}{lclcccc}
  Model  & $\chi^2_{\rm min}/N_d$ & $\varepsilon_1$(50 MeV)
         & $P_{\rm d}$(\%)        & $E_{\rm B}$(MeV)
         & $\langle V_T \rangle$  & $\langle V_{\pi} \rangle$     \\
\tableline
Nijm PWA93& 0.99 & 2.11$\pm$0.05 & --   & --   &  --   &  --      \\
 & & & & & & \\
 Nijm~I   & 1.03 & 2.09          & 5.66 & 7.72 &  72\% &  74\%    \\
 Nijm92   & 1.92 & 1.98          & 5.61 & 7.68 &  77\% & 113\%    \\
 & & & & & & \\
 Nijm~II  & 1.03 & 2.00          & 5.64 & 7.62 &  52\% &  67\%    \\
 Reid93   & 1.03 & 2.03          & 5.70 & 7.63 &  57\% &  71\%    \\
 AV18     & 1.30 & 2.16          & 5.65 & 7.62 &  52\% &  77\%    \\
\end{tabular}
\label{results}
\end{table}

\begin{figure}
\caption{The $\varepsilon_1$ mixing parameter in degrees as a
         function of $T_{\rm lab}$ in MeV. The shaded band
         represents the multi-energy solution with its statistical
         multi-energy error. The black dots represent the
         single-energy determinations with their single-energy
         errors.}
\label{eps1}
\end{figure}


\begin{references}
\bibitem{Fad60} L.D.\ Faddeev, Zh.\ Eksp.\ Teor.\ Fiz.\ {\bf 39},
         1459 (1960) [Sov.\ Phys.\ - JETP {\bf 12}, 1014 (1961)].
\bibitem{Che85} C.R.\ Chen, G.L.\ Payne, J.L.\ Friar, and
         B.F.\ Gibson, Phys.\ Rev.\ C {\bf 31}, 2266 (1985);
         {\bf 33}, 1740 (1986);
         J.L.\ Friar, B.F.\ Gibson, and G.L.\ Payne,
         {\it ibid}, {\bf 35}, 1502 (1987);
         J.L.\ Friar, B.F.\ Gibson, and G.L.\ Payne,
         {\it ibid}, {\bf 37}, 2869 (1988); these references
         contain a representative set of potential model results.
\bibitem{Car87} J.\ Carlson, Phys.\ Rev.\ C {\bf 36}, 2026 (1987).
\bibitem{Carunp} J. Carlson, (to be published).
\bibitem{Che91} C.R.\ Chen, G.L.\ Payne, J.L.\ Friar, and
         B.F.\ Gibson,
         Phys.\ Rev.\ C {\bf 44}, 50 (1991).
\bibitem{Fri91} J.L.\ Friar, B.F.\ Gibson, H.C.\ Jean, and
         G.L.\ Payne,
         Phys.\ Rev.\ Lett.\ {\bf 66}, 1827 (1991).
\bibitem{Glo90} W.\ Gl\"ockle, H.\ Wita{\l}a, and Th.\ Cornelius,
         Nucl.\ Phys.\ {\bf A508}, 115c (1990);
         W.\ Gl\"ockle and H.\ Wita{\l}a,
         Kerntechnik {\bf 57}, 228 (1992).
\bibitem{Wu90} Y.\ Wu, S.\ Ishikawa, and T.\ Sasakawa,
         Phys.\ Rev.\ Lett.\ {\bf 64}, 1875 (1990);
         {\bf 66}, 242(E) (1991).
\bibitem{Coo79} S.A.\ Coon, M.D.\ Scadron, P.C.\ McNamee,
         B.R.\ Barrett, D.W.E.\ Blatt, and B.H.J.\ McKellar,
         Nucl.\ Phys.\ {\bf A317}, 242 (1979); S.A.\ Coon and
         J.L.\ Friar, Phys.\ Rev.\ C {\bf 34}, 1060 (1986).
\bibitem{Car93} J.\ Carlson, V.R.\ Pandharipande, and R.\ Schiavilla,
         Phys.\ Rev.\ C {\bf 47}, 484 (1993).
\bibitem{Rup92} G.\ Rupp and J.A.\ Tjon,
         Phys.\ Rev.\ C {\bf 45}, 2133 (1992).
\bibitem{Glo86} W.\ Gl\"ockle, T.-S.\ H.\ Lee, and F.\ Coester,
         Phys.\ Rev.\ C {\bf 33}, 709 (1986).
\bibitem{Sam92} F.\ Sammarruca, D.\ P. Xu, and R.\ Machleidt,
         Phys.\ Rev.\ C {\bf 46}, 1636 (1992).
\bibitem{St93a} V.\ Stoks and J.J.\ de Swart,
         Phys.\ Rev.\ C {\bf 47}, 761 (1993).
\bibitem{Fri87} J.L.\ Friar, B.F.\ Gibson, and G.L.\ Payne,
         Phys.\ Rev.\ C {\bf 36}, 1140 (1987).
\bibitem{Ber88} J.R.\ Bergervoet, P.C.\ van Campen,
         W.A.\ van der Sanden, and J.J.\ de Swart,
         Phys.\ Rev.\ C {\bf 38}, 15 (1988).
\bibitem{Ber90} J.R.\ Bergervoet, P.C.\ van Campen,
         R.A.M.\ Klomp, J.-L.\ de Kok, T.A.\ Rijken,
         V.G.J.\ Stoks, and J.J.\ de Swart,
         Phys.\ Rev.\ C {\bf 41}, 1435 (1990).
\bibitem{Klo92} R.A.M.\ Klomp, T.A.\ Rijken, V.G.J.\ Stoks, and
         J.J.\ de Swart,
         Few-Body Systems, Suppl.\ {\bf 6}, 105 (1992).
\bibitem{St93b} V.G.J.\ Stoks, R.A.M.\ Klomp, M.C.M.\ Rentmeester,
         and J.J.\ de Swart, in preparation.
\bibitem{Klo91} R.A.M.\ Klomp, V.G.J.\ Stoks, and J.J.\ de Swart,
         Phys.\ Rev.\ C {\bf 44}, R1258 (1991).
\bibitem{Ham91} M.\ Hammans, C.\ Brogli-Gysin, S.\ Burzynski,
         J.\ Campbell, P.\ Haffter, R.\ Henneck, W.\ Lorenzon,
         M.A.\ Pickar, I.\ Sick, J.A.\ Konter, S.\ Mango, and
         B.\ van den Brandt,
         Phys.\ Rev.\ Lett.\ {\bf 66}, 2293 (1991).
\bibitem{Haf92} P.\ Haffter, C.\ Brogli-Gysin, J.\ Campbell,
         D.\ Fritschi, J.\ G\"otz, M.\ Hammans, R.\ Henneck,
         J.\ Jourdan, G.\ Masson, L.M.\ Qin, S.\ Robinson,
         M.\ Tuccillo, J.A.\ Konter, S.\ Mango, and
         B.\ van den Brandt, Nucl.\ Phys.\ {\bf A548}, 29 (1992).
\bibitem{Nag78} M.M.\ Nagels, T.A.\ Rijken, and J.J.\ de Swart,
         Phys.\ Rev.\ D {\bf 17}, 768 (1978).
\bibitem{RSC68} R.V.\ Reid,
         Ann.\ Phys.\ (N.\ Y.\ ) {\bf 50}, 411 (1968).
\bibitem{Wir84} R.B.\ Wiringa, R.A.\ Smith, and T.A.\ Ainsworth,
         Phys.\ Rev.\ C {\bf 29}, 1207 (1984).
\bibitem{St93c} V.G.J.\ Stoks, R.A.M.\ Klomp, and J.J.\ de Swart,
         in preparation.
\bibitem{Cha77} K.\ Chadan and P.\ C.\ Sabatier,
        {\it Inverse Problems in Quantum Scattering Theory},
        (Springer-Verlag, New York, 1977).
\end{references}
\end{document}